\documentclass[aps,pre,twocolumn,floatfix,nofootinbib]{revtex4}
\usepackage{graphicx,epsf}
\usepackage{amsmath,amsfonts} % to use \mathbb

\begin{document}

\newcommand{\EQ}{Eq.~}
\newcommand{\EQS}{Eqs.~}
\newcommand{\FIG}{Fig.~}
\newcommand{\FIGS}{Figs.~}
\newcommand{\SEC}{Sec.~}
\newcommand{\SECS}{Secs.~}

\title{Extremal dynamics on complex networks: Analytic solutions}

\author{N. Masuda}
\affiliation{Laboratory for Mathematical Neuroscience, RIKEN Brain
Science Institute, 2-1, Hirosawa, Wako, Saitama 351-0198, Japan}
\author{K.-I. Goh}
\author{B. Kahng}
\affiliation{School of Physics and Center for Theoretical Physics,
Seoul National University, 151-747, Korea}
\date{\today}

\begin{abstract}
The Bak-Sneppen model displaying punctuated equilibria in
biological evolution is studied on random complex networks. By
using the rate equation and the random walk approaches, we obtain
the analytic solution of the fitness threshold $x_c$ 
to be $1/(\langle k \rangle_f+1)$, where 
$\langle k \rangle_f=\langle k^2 \rangle/\langle k \rangle$ 
($=\langle k \rangle$) in the quenched (annealed) updating case,  
where $\langle k^n \rangle$ is the $n$-th moment of the degree
distribution. Thus, the threshold is zero (finite) for the degree
exponent $\gamma <3$ ($\gamma > 3$) for the quenched case 
in the thermodynamic limit.
The theoretical value $x_c$ fits well to the numerical simulation 
data in the annealed case only.
Avalanche size, defined as the duration of successive mutations below
the threshold, exhibits a critical behavior as its distribution 
follows a power law, $P_a(s)\sim s^{-3/2}$. 
\end{abstract}

% 89.75.Hc networks and genealogical trees
% 89.75.Da systems obeying scaling laws
% 89.75.Fb structures and organization in complex systems
\maketitle

\section{Introduction}\label{sec:introduction}
Punctuated equilibrium is an evolution taking place through
intermittent bursts of activity separating relatively long periods
of quiescence, which can be often found in ecological
systems~\cite{gould,raup}. Bak and Sneppen (BS)~\cite{Bak93}
introduced a simple model to mimic such an evolution. The
basis of the BS model is to focus on a minimal set of
variables that capture the basic features of punctuated
equilibrium while ignoring all other details. In the original
model, $N$ species are arranged on a one-dimensional chain with
periodic boundary conditions. A fitness value $x_i$ is assigned to
each site $i$ (species) on the chain, which is a random variable
selected in the interval [0,1]. Evolution in the ecological
systems is modeled as follows: At each time step, the ecological
system is updated by locating the site with the lowest fitness and
mutating it by assigning new random numbers to that site and the
$K-1$ nearest neighboring sites. Subsequent updating of the lowest
fitness value generates spatial and temporal correlations and
displays punctuated equilibria. A distinct feature arising through
this dynamics can be found in the distribution of fitness values.
After a transient period, the distribution of the fitness values
has a discontinuity at a threshold $x_c\approx 0.67$; its elements
are zero up to $x_c$ and almost the same constant beyond $x_c$.
The threshold $x_c$ is self-organized.

A mean field version of the BS model was
introduced~\cite{Flyvbjerg}, in which updated are the minimum
fitness value as well as the fitness values of other $K-1$ sites
selected at random in the system. Such a modified model enables
one to solve the problem analytically. The threshold was obtained
to be $x_c=1/K$ in the limit $N \to \infty$. Also the notion of
avalanche was introduced to quantify the correlation between
bursts of evolutionary activity. Avalanche size is the time
interval between two successive occasions where no fitness value
is less than a given value. It was proposed based on the branching
process analysis~\cite{harris} and later derived by using the
random walk approach~\cite{Deboer94,Deboer95} that the avalanche
size distribution follows a power law as $P_a (s) \sim s^{-3/2}$
when the given value is chosen as the threshold $x_c$.

Ecological systems in real world are complex. Interactions between
individual species are not as simple as one-dimensional, but form
a complex network. Thus, it would be interesting to extend
previous studies of the BS model performed in the Euclidean space to
complex networks such as scale free (SF) networks, while it is
still controversial whether the ecological systems such as food webs
are SF-networked systems~\cite{food_web}. SF networks mean that the
number of connections to each species, called degree $k$ in graph
theory, follows a power law, $P_d(k)\sim
k^{-\gamma}$~\cite{review1,review2,review3,review4}. While such an
extension is natural, only few studies in that direction have
been performed so far. Christensen $et$ $al.$~\cite{Christensen}
have studied the BS model on random networks~\cite{er}. Kulkarni $et$
$al.$~\cite{kulkarni} studied it on the small-world network
introduced by Watts and Strogatz~\cite{ws}. Moreno
and Vazquez~\cite{moreno} studied the BS model on SF networks with
$\gamma=3$, obtaining that the threshold $x_c$ is given as
$x_c=\langle k \rangle/\langle k^2 \rangle$ by using the heuristic
argument similar to the one used in the contact process, where
$\langle \cdots \rangle$ denotes the average over the degree
distribution. They found that $x_c$ depends on the system size $N$ as
$x_c \sim 1/\ln N$, so that it vanishes in the limit $N\to
\infty$. The $N$-dependent behavior was obtained numerically at
$\gamma=3$, so that the result may be rooted from logarithmic
correction. Recently, Lee and Kim~\cite{ykim} also studied the
same problem on SF networks but with general $\gamma > 2$. They
obtained that the threshold is given as $x_c=(\langle k
\rangle+1)/\langle (k+1)^2 \rangle$ by using heuristic arguments.
Thus the threshold vanishes for $\gamma < 3$ and finite for
$\gamma > 3$. The interesting feature they obtained is the
crossover behavior in the avalanche size distribution between two
different power-law behaviors.

Here we study the BS model on random SF networks analytically by
using both the rate equation and the random walk
approaches~\cite{Rednerbook}. By random SF networks, we mean the
SF network with no degree-degree correlation. The rate equation is
set up for the case that updating of the fitness values is carried
out not only at the vertex with the smallest fitness value but
also at its nearest neighbors, which is called quenched 
case. We also compare the quenched case with the annealed case, 
where updating is carried out at the vertex with the minimum 
fitness values as well as the vertices randomly chosen
over the entire system and its number is equal to the degree of the 
vertex with the minimum fitness value. It is noteworthy that 
the number of vertices updated is not constant in complex networks, 
but depends on the degree of the vertex with the minimum fitness value. 
Thus, the analytic approach of the BS model is not as simple as 
the case in the Euclidean space.
Here by applying the rate equation approach as well as the random
walk approach, we obtain the fitness threshold analytically to be
$x_c=1/(\langle k \rangle_f+1)$, where 
$\langle k \rangle_f=\langle k^2 \rangle/\langle k \rangle$ in 
the quenched case and $\langle k \rangle_f=\langle k \rangle$ in
the annealed case.
The avalanche size distribution is obtained to be 
$P_a(s)\sim s^{-\tau}$ with $\tau=3/2$ at $x_c$ for $\gamma > 3$.

\section{Rate equation approach}

In SF networks, it would be essential to take it into account 
the fact that vertices with different degrees experience different
updating frequencies. Let us denote by $f_k$ the probability that 
the vertex
with the smallest fitness value has degree $k$. $\rho_k(x)$ is the
distribution function of fitness values at the vertices with
degree $k$.
Here we set up the master equation for the fitness distribution
for the quenched updating, following \cite{Flyvbjerg}. First we define a
quantity $Q_k(x)$, the accumulative distribution of the fitness
values at vertices with degree $k$:
\begin{equation}
Q_k(x)=\int^1_{x} dx^{\prime} \rho_k(x^{\prime}).
\end{equation}
Thus the fitness distribution $\rho_k(x)=-\frac{\partial}{\partial
x}Q_k(x)$. Then, we have
\begin{equation}
f_k=-\int^1_0 dx^{\prime} \frac{\partial}{\partial x^{\prime}}
\left\{ Q_k(x^{\prime})\right\}^{N p_k} \prod_{k{\prime}\neq k}
\left\{ Q_{k^{\prime}}(x^{\prime})\right\}^{N p_{k^{\prime}}},
\end{equation}
where $p_k$ denotes the degree distribution $P_d(k)$ for
simplicity. Note that $\sum_k f_k = 1$ is satisfied. The evolution
equation for the fitness distribution at vertices with degree $k$
is written as
\begin{widetext}
\begin{align}
\rho_k(x,t+1)= & \,\,\rho_k(x,t) -\frac{f_k}{Np_k} \left[
\frac{-\frac{\partial}{\partial x}\left\{Q_k(x)\right\}^{N p_k}
\prod_{k^{\prime}\neq k}\left\{Q_{k^{\prime}}(x)\right\}^{N
p_{k^{\prime}}} }
{f_k}\right]\nonumber\\
&- \sum_{k^{\prime\prime}}k^{\prime\prime}f_{k^{\prime\prime}}
\frac{k p_k}{\left<k\right>} \left[\frac{\rho_k(x,t)-\frac{f_k}{N
p_k}\left[-\frac{\partial}{\partial x} \left\{ Q_k(x)\right\}^{N
p_k} \prod_{k^{\prime}\neq k}\left\{Q_{k^{\prime}}(x)\right\}^{N
p_{k^{\prime}}}\cdot\frac{1}{f_k}\right]}{n p_k - f_k}
\right]\nonumber \\
&+ \frac{f_k}{Np_k}+\sum_{k^{\prime\prime}} \frac{k
p_k}{\left<k\right>}\frac{f_{k^{\prime\prime}}k^{\prime\prime}}{Np_k},
\label{master}
\end{align}
where the second term of the right-hand side (RHS) of
Eq.(\ref{master}) represents the update of the minimum fitness
when it locates at the vertex with degree $k$. The third term does
the update of the fitness value of a vertex with degree $k$
induced by a nearest neighboring vertex with degree
$k^{\prime\prime}$ which has the minimum fitness value in the
system. The factor $kp_k /\langle k \rangle$ comes from the
conditional probability $P(k|k^{\prime\prime})$ that the vertex
with degree $k$ is connected to the one with $k^{\prime\prime}$, 
which is relevant to the quenched case. In the annealed 
case, the factor is replaced with $p_k$ simply.
The last two terms do the addition of new fitness
values~\cite{Flyvbjerg}.

The stationary solution in the limit $t\to\infty$ can be solved by
using $\rho_k(x)=-\frac{\partial}{\partial x}Q_k(x)$ and
taking the integral over $[x,1]$ of the whole formula as the
integral equation,
\begin{align}
- \int^1_x dx^{\prime} \frac{\partial}{\partial x^{\prime}}
\left\{Q_k(x^{\prime})\right\}^{N p_k} \prod_{k^{\prime}\neq k} &
\left\{ Q_{k^{\prime}}(x^{\prime})\right\}^{N p_{k^{\prime}}}
\left[ -\frac{1}{N p_k} + \frac{k\left<k\right>_f}
{\left<k\right>\left(N p_k - f_k\right)N}\right]\nonumber \\
&-\frac{\left<k\right>_f k p_k}{\left<k\right>\left(N p_k - f_k
\right)}Q_k(x) + \frac{f_k+\frac{k p_k
\left<k\right>_f}{\left<k\right>}}{N p_k} (1-x) = 0,
\label{eq:Q_k-basic}
\end{align}
\end{widetext}
where
\begin{equation} \left<k\right> = \sum^{\infty}_{k=1} k
p_k\quad \hbox{and} \quad \left<k\right>_f = \sum^{\infty}_{k=1} k
f_k.
\end{equation}

As done in \cite{Flyvbjerg}, the threshold $x_c$ is determined by
the comparison of the first term with the second term of Eq.
(\ref{eq:Q_k-basic}) in their absolute magnitudes. To proceed, let
us first assume that the second term is dominant compared with the
first term. Then, we obtain that within the leading order,
\begin{equation}
Q_k(x) \approx \left(\frac{\left<k\right>f_k}{\left<k\right>_f k
p_k} + 1 \right)(1-x), \label{eq:Q_k}
\end{equation}
which leads to
\begin{equation}
\rho_k(x) \cong \frac{\left<k\right> f_k}{\left<k\right>_f k
p_k}+1. \label{eq:rho_k}
\end{equation}
This result holds when $Q_k(x)$ is less than 1 by more than
$\mathcal{O}(1/N)$. In other words,
\begin{equation}
x \gg x_{c,k}+\mathcal{O}(1/N) =
\frac{\left<k\right>f_k}{\left<k\right>f_k+\left<k\right>_f k p_k}
+\mathcal{O}(1/N), \label{eq:x_c}
\end{equation}
where $x_{c,k}$ is the threshold for a given $k$.
\EQ(\ref{eq:rho_k}) indicates that $\rho_k(x)$ does not vanish as
$N\to\infty$. We show that $x_{c,k}$ does not depend on $k$ in
Appendix. Thus we denote $x_{c,k}=x_c$ simply.

For $x<x_c$, the second term of \EQ(\ref{eq:Q_k-basic}) can be
ignored and we have
\begin{align}
\int^1_{x} dx^{\prime} \frac{N p_{k} \rho_{k}(x^{\prime})}
{Q_{k}(x^{\prime})} \prod_{k^{\prime}} & \left\{
Q_{k^{\prime}}(x^{\prime})\right\}^{N p_{k^{\prime}}} \left[ 1 -
\frac{k\left<k\right>_f} {\left<k\right>N}\right] \nonumber \\ & \cong \left(f_k +
\frac{k p_k\left<k\right>_f} {\left<k\right>} \right)(1-x).
\label{eq:x-small}
\end{align}

Next we use the fact that the integral $\int^1_{x_c}\cdots$ is
very small and the interval of the integration is replaced by
$\int_{x}^{x_c}$. 
By setting $x=0$, we have
\begin{equation}
n p_k\rho_k \propto f_k. \label{relation1}
\end{equation}
On the other hand, by differentiating \EQ(\ref{eq:x-small}) at
$x=0$, we obtain
\begin{equation}
n p_k \rho_k = f_k + \frac{k p_k
\left<k\right>_f}{\left<k\right>}.
\label{relation2}
\end{equation}
Combining Eqs.(\ref{relation1}) and (\ref{relation2}), we obtain
that
\begin{equation}
f_k = \frac{k p_k}{\left<k\right>}, \label{f_k}
\end{equation}
which is supported by numerical simulations shown in Fig.~1. Note 
that the result of Eq.(\ref{f_k}) is for the quenched case. 
In the annealed case, similar calculations lead to $f_k=p_k$. 
Thus we obtain 
$\left<k\right>_{f}=\left<k\right>_{f,q}=\left<k^2\right>/\langle k \rangle$ 
in the quenched case and $\left< k \right>_f=\left<k\right>_{f,a}=\left < k \right>$ 
in the annealed case.  $\langle k \rangle_{f,q}$ diverges for $\gamma<3$
as $\sim N^{(3-\gamma)/(\gamma-1)}$ in the limit $N\to \infty$.

\begin{figure}
\centerline{\epsfxsize=8.5cm \epsfbox{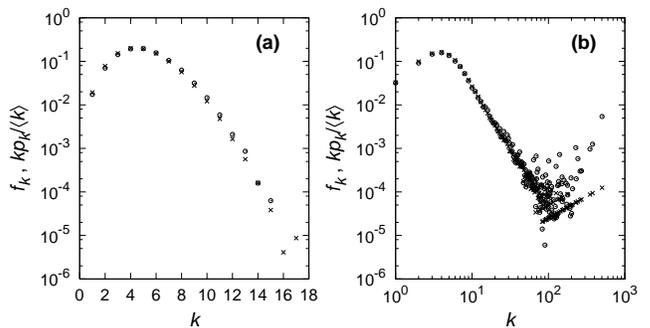}} \caption{
The data of $f_k$
$(\circ)$ and $kp_k/\langle k\rangle$ $(\times)$ for the
Erd\H{o}s-R\'enyi network (a) and for the scale-free network with
$\gamma=3.6$ (b) in the quenched cases. 
To generate the scale-free network, we use the
static model \cite{static}. Both networks have the average degree $\langle
k\rangle=4$ and the system size $N=10^6$.}
\end{figure}
We also have
\begin{equation}
\rho_k=\frac{k\left(\left<k\right>_f+1\right)}{n\left<k\right>}.
\end{equation}
Thus, $\rho_k \sim \mathcal{O}(N^{-(\gamma-2)/(\gamma-1)})$, and
converges to 0 as $N\to\infty$. The convergence rate is slower
than the rate of $1/N$ that appears in Euclidean space. Finally, from
\EQ(\ref{eq:x_c}), the threshold can be expressed simply as 
\begin{equation}
x_c = \frac{1}{\left<k\right>_f+1},
\label{x_c}
\end{equation}
which does not depend on $k$ for both updating rules. 
This result is different 
from the previous results~\cite{moreno,ykim}. The threshold formula is
reproduced by the random walk approach in the next section.

\section{Random walk approach}

The random walk approach was first introduced in
\cite{Deboer94,Deboer95}, and it is useful for calculating the
avalanche size distribution. The threshold can be also obtained.
Let $q_{\lambda}(t)$ be the probability of having an avalanche 
with size $t$, which is defined as the duration of time
throughout which the minimum fitness value is smaller than a
given threshold value $\lambda$.
$\lambda$ can be chosen arbitrary. Later we
find that $q_{\lambda}(t)$ follows a power law when $\lambda$ is
equal to the threshold $x_c$. The corresponding generating
function is defined as $ \chi(z)=\sum_{t>0} q_{\lambda}(t) z^t$.
Then $\chi(z)$ satisfies the self-consistent
equation~\cite{Deboer95,harris,Goh03prl},
\begin{equation}
\chi(z)=z g(\chi(z)). \label{eq:chi-self}
\end{equation}
In the previous study~\cite{Deboer95}, the generating function
$g(z)$ was given as $\sum_t {K\choose t}\lambda^t
(1-\lambda)^{K-t}z^t=\left(1-\lambda + \lambda z\right)^K$ when
$K$ fitness values are updated randomly. However, in the case of
SF networks, the number of vertices updated at each time step is
not constant, but it depends on the degree of the vertex with the
minimum fitness value. In this case, the generating function
$g(z)$ is given as
\begin{equation}
g(z)=\sum_{k=1}^{\infty}f_{k-1}\left(1-\lambda + \lambda
z\right)^k, \label{g_z}
\end{equation}
where $f_k$ was defined as the probability that the minimum
fitness locates at the vertex with degree $k$.

What we do next is to solve $q_{\lambda}(t)$ by using
Eqs.(\ref{eq:chi-self},\ref{g_z}). To proceed, we use the
Lagrange's inversion formula~\cite{Grimmettbook,masuda},
\begin{equation}
h\left(\omega\right)=h(0)+\sum^{\infty}_{n=1}\frac{z^n}{n!} \left[
\frac{d^{n-1}}{du^{n-1}} \left[h^{\prime}(u)g(u)^n\right]
\right]_{u=0}, \label{eq:Lagrange}
\end{equation}
where $z=\omega/g(\omega)$, provided that $\omega/g(\omega)$ is
analytic near $\omega=0$ and $h(\omega)$ is an infinitely
differentiable function. Here we choose $h(\omega)=\omega$ and
$\omega(z)=\chi(z)$. Then,
\begin{widetext}
\begin{eqnarray}
\chi(z) &=& \sum^{\infty}_{t=1} \frac{z^t}{t!}
\frac{\partial^{t-1}}{\partial u^{t-1}}
\left[\left(g(u) \right)^t \right]_{u=0}\nonumber\\
&=& \sum^{\infty}_{t=1} \frac{z^t}{t!}
\frac{\partial^{t-1}}{\partial u^{t-1}} \left[ \prod_{i=1}^t
\sum_{k_i=1}^{\infty} f_{k_i-1} \left(1-\lambda + \lambda
u\right)^{k_i}
\right]_{u=0}\nonumber\\
&=& \sum^{\infty}_{t=1} \frac{z^t}{t!}
\frac{\partial^{t-1}}{\partial u^{t-1}} \left[\sum^{\infty}_{k_1=1}
\ldots \sum^{\infty}_{k_t=1} \left(\prod_{i=1}^t f_{k_i-1}\right)
\left(1-\lambda + \lambda u\right)^{\sum^t_{i=1}k_i}
\right]_{u=0}\nonumber\\
&=& \sum^{\infty}_{t=1} \frac{z^t}{t!} \sum^{\infty}_{k_1=1}
\ldots \sum^{\infty}_{k_t=1} \left(\prod_{i=1}^t f_{k_i-1}\right)
\frac{(\sum^t_{i=1}k_i)!}{(\sum^t_{i=1}k_i-t+1)!}
\lambda^{t-1}\left(1-\lambda \right)^{\sum^t_{i=1}k_i-t+1}.
\end{eqnarray}
Since $\chi(z)=\sum_t q_{\lambda}(t)z^t$, $q_{\lambda}(t)$ can be
obtained by using the Stirling's formula as
\begin{eqnarray}
q_{\lambda}(t) &=& \frac{1}{t!} \sum^{\infty}_{k_1=1} \ldots
\sum^{\infty}_{k_t=1} \left(\prod_{i=1}^t f_{k_i-1}\right)
\frac{(\sum^t_{i=1}k_i)!}{(\sum^t_{i=1}k_i-t+1)!}
\lambda^{t-1}\left(1-\lambda \right)^{\sum^t_{i=1}k_i-t+1}\nonumber\\
&=& \sum^{\infty}_{k_1=1} \ldots \sum^{\infty}_{k_t=1}
\left(\prod_{i=1}^t f_{k_i-1}\right)
\frac{(\sum^t_{i=1}k_i)^{\sum^t_{i=1}k_i}}
{(\sum^t_{i=1}k_i-t)^{\sum^t_{i=1}k_i-t} \; t^t}\cdot
\frac{1}{\sum^t_{i=1}k_i-t+1}\cdot\nonumber\\
&&\frac{\sqrt{2\pi \sum^t_{i=1}k_i}} {\sqrt{2\pi
\left(\sum^t_{i=1}k_i-t\right)} \sqrt{2\pi t}}
\lambda^{t-1}\left(1-\lambda \right)^{\sum^t_{i=1}k_i-t+1}.
\label{eq:q(t)-1}
\end{eqnarray}
Note that $k_i$ is the degree of the vertex
with the minimum fitness value at updating time $i$ plus 1, which
occurs with the probability $f_{k_i-1}$. Thus in the limit $t\to
\infty$
\begin{equation}
\sum^t_{i=1}k_i \cong t \cdot \langle k + 1 \rangle_f =
\left(\langle k \rangle_f+1\right) t. \label{eq:q(t)-2}
\end{equation}
Substituting \EQ(\ref{eq:q(t)-2}) into \EQ(\ref{eq:q(t)-1}) yields
\begin{eqnarray}
q_{\lambda}(t) &\cong& \frac{1-\lambda}{\lambda}
\sqrt{\frac{\left<k\right>_f+1}{2\pi \left<k\right>_f^3}}
\left[\frac{\lambda (1-\lambda)^{\left<k\right>_f}
(\left<k\right>_f+1)^{\left<k\right>_f+1}}
{(\left<k\right>_f)^{\left<k\right>_f}} \right]^t
t^{-3/2}+\mathcal{O}(t^{-5/2}). \label{eq:q(t)-3}
\end{eqnarray}

\end{widetext}
The quantity in the square bracket is 1 when
\begin{equation}
\lambda=\frac{1}{\left<k\right>_f+1},  
\label{eq:lambda_c-avalanche}
\end{equation}
which is equal to the threshold $x_c$ previously obtained via the
rate equation approach. Then, $q_{x_c}(t)\sim t^{-3/2}$.
Therefore, the avalanche size distribution behaves as $P_a (s)
\sim s^{-3/2}$.

\section{Numerical results}

\begin{figure}
\centerline{\epsfxsize=8.5cm \epsfbox{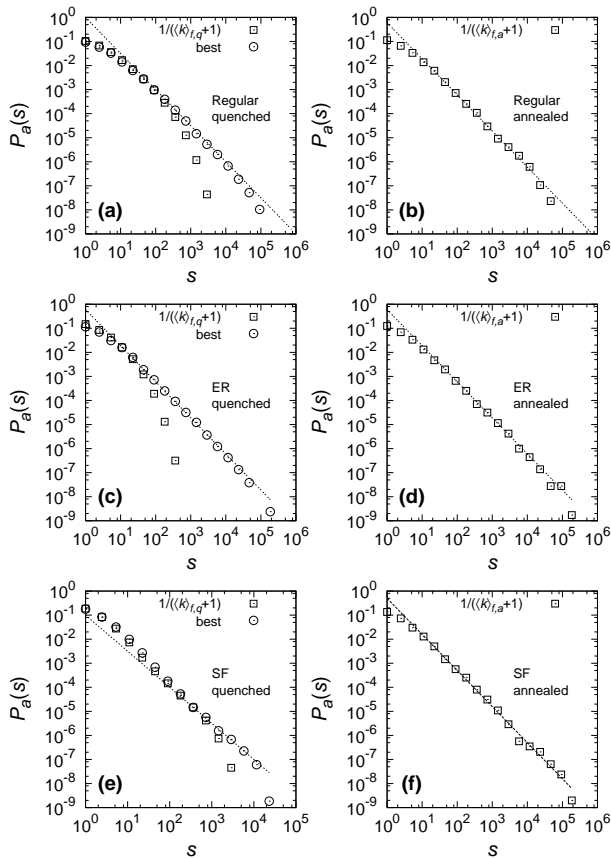}} \caption{
(a) and (b). The avalanche size distribution for 
the regular network with the degree distribution 
of the $\delta$-function.
In the quenched case (a), the avalanche
size distribution follows a power law when $\lambda$ is chosen as
0.253 ($\circ$), larger than the theoretical value $1/(\langle k
\rangle+1)=0.2$ ($\square$). In the annealed case (b), the
theoretical value $x_c=0.2$ ($\circ$) works well to generate the
power-law behavior of $P_a(s)$. 
(c) and (d). Same plot for the  
Erd\H{o}s-R\'enyi network \cite{er}. In the quenched case (c), 
the power-law behavior of $P_a(s)$ occurs at $x_c=0.207$, 
larger than the theoretical value, $x_c=1/6$. In the annealed 
case (d), the theoretical value $x_c$=0.2 generates the power-law 
behavior. 
(e) and (f). Same plot for the scale-free network with $\gamma=3.6$.
In the quenched case (e), the power-law behavior of 
$P_a(s)$ occurs at $x_c=0.1$, which is larger than the 
theoretical value $x_c=0.08$. In the annealed case, the 
theoretical value $x_c=0.2$ yields the power-law 
behavior of $P_a(s)$. 
The mean degree is fixed to be $\langle k \rangle=4$ and
the system size is $N=10^6$ in all cases. 
The straight lines have slope -3/2 in all cases.} \label{test_random}
\end{figure}

We check the analytical solution of $q_{\lambda}(t)$ numerically
for several networks. The theoretical formula of $x_c$ is tested
through the criticality of the avalanche size distribution.
First, the random network with the degree distribution
$P_d(k)=\delta_{k,k_0}$, called the regular network, is constructed 
and the dynamics of the BS model is performed on that network.  
$k_0=4$ is taken for numerical simulations. 
In this case, $x_c$ reduces to $x_c=1/(k_0+1)$ simply in both 
the quenched and annealed cases. 
In the quenched case (Fig.~\ref{test_random}(a)), 
the avalanche size distribution does not follow a power law 
when $\lambda=x_c=0.2$. Instead, the power-law behavior appears 
at a larger value, $\lambda \approx 0.253$. 
In the annealed case (Fig.~\ref{test_random}(b)), 
however, it follows a power law at $\lambda=x_c$, 
consistent with the theoretical value. 

Second, for Erd\H{o}s-R\'enyi (ER) random graph, where the degree 
distribution is a Poisson distribution, the theoretical 
formula reduces to $x_c \approx 1/(\langle k \rangle+2)$ 
in the quenched case, because $\langle k^2 \rangle =
\langle k \rangle^2+\langle k \rangle$ in the limit $N\to \infty$. 
Numerical simulations are performed
in both the quenched and the annealed cases. 
In the quenched case (Fig.~\ref{test_random}(c)), 
the avalanche size distribution does not follow a power law when
$\lambda$ is taken as the theoretical value, but it does 
when $\lambda \approx 0.207$. 
In the annealed case (Fig.~\ref{test_random}(d)), 
the avalanche size distribution follows a power law
at $\lambda=1/(\langle k \rangle+1)$, consistent with 
the theoretical value. 

Next, for SF networks with $\gamma=3.6$, which is constructed 
through the static model~\cite{static}, the theoretical value 
$x_c=1/(\langle k \rangle_f +1)\approx 0.08$ in the quenched 
case. Note that $\langle k \rangle_f \approx 11.5$ is different 
from $\langle k^2 \rangle /\langle k \rangle \approx 7.06$ 
numerically due to the strong fluctuations arising in the large 
$k$ region (Fig.~1). Again the avalanche size distribution does 
not follow a power law at the theoretical value, but does at 
$\lambda \approx 0.1$. In the annealed case, the avalanche 
size distribution follows a power law at
$x_c=1/(\langle k \rangle_f+1)$. 

The numerical results for the above three networks indicate 
that the mean-field theoretical prediction is not good for the 
quenched case, however, is good for the annealed case 
instead. This result is attributed to the effect of the temporal 
and spatial correlation between the vertices with
the minimum fitness value at successive time steps, which often
occur at the nearest neighbors or at the same vertex. 
Such effect was not counted properly in the quenched case, 
and can be neglected in the annealed case. 

\section{Conclusions}
We have studied the Bak-Sneppen model on complex networks by using
the master equation as well as the random walk approaches. The
threshold $x_c$ is obtained to be $x_c=1/(\langle
k\rangle_f+1)$, where $\langle \cdots \rangle_f$
is the average over the minimum fitness vertices. The avalanche size
distribution follows a power law with the exponent $\tau=3/2$ at
the critical point. The theoretical prediction of $x_c$ was tested
numerically for the regular network, the ER random network, 
and the SF network. For all the networks, the theoretical 
predictions of $x_c$ are in disagreement (agreement) with the 
numerical results for the quenched (annealed) case. 
The discrepancy in the quenched case is attributed to the effect 
of the temporal and spatial correlation between the vertices with 
the minimum fitness values at successive time steps. Nevertheless, 
the formula $x_c=\langle k \rangle/(\langle k^2 \rangle+\langle k \rangle)$ 
is newly derived here for the quenched case. Thus when 
$2 < \gamma < 3$, $x_c\to 0$ in the thermodynamic limit in 
the quenched case.  

\appendix
\section{}

Here we show that $x_{c,k}$ does not depend on $k$. To proceed, we
suppose $x_{k_1,c} < x_{k_2,c}$ for a certain pair of $k_1$ and
$k_2$. Then, for a given $x_0$ in the range $x_{k_1,c}< x_0
<x_{k_2,c}$, we have with Eq.~(\ref{eq:Q_k-basic}),
\begin{align}
\int^1_{x_0} dx^{\prime} \frac{N p_{k_1} \rho_{k_1}(x^{\prime})}
{Q_{k_1}(x^{\prime})} \prod_{k^{\prime}} & \left\{
Q_{k^{\prime}}(x^{\prime})\right\}^{N p_{k^{\prime}}} \left[ 1 -
\frac{k_1\left<k\right>_f} {\left<k\right>N}\right] \nonumber \\ & \ll
\frac{\left<k\right>_f k_1 p_{k_1}}{\left<k\right>}Q_{k_1}(x_0),
\label{eq:compare-k_1}
\end{align}
and
\begin{align}
\int^1_{x_0} dx^{\prime} \frac{N p_{k_2} \rho_{k_2}(x^{\prime})}
{Q_{k_2}(x^{\prime})} \prod_{k^{\prime}} & \left\{
Q_{k^{\prime}}(x^{\prime})\right\}^{N p_{k^{\prime}}} \left[ 1 -
\frac{k_2\left<k\right>_f} {\left<k\right>N}\right] \nonumber \\
& \cong
\frac{\left<k\right>_f k_2 p_{k_2}}{\left<k\right>}Q_{k_2}(x_0).
\label{eq:compare-k_2}
\end{align}

Since $Q_{k_1}(x)$ and $Q_{k_2}(x)$ are of the same order based on
\EQ(\ref{eq:Q_k}), the RHS of \EQ(\ref{eq:compare-k_1}) and that
of \EQ(\ref{eq:compare-k_2}) are of the same order. Then we must
have $\rho_{k_1}(x^{\prime}) \ll \rho_{k_2}(x^{\prime})$ for
$x_0\le x^{\prime}\le 1$, which contradicts \EQ(\ref{eq:rho_k})
for $x>x_{k_2,c} (>x_0, x_{k_1,c})$. Thus we set $x_c = x_{k,c}$
for all $k$.\\

\begin{acknowledgments}
This work is supported in part by special postdoctoral researchers
program of RIKEN and in part by the KRF Grant funded by the Korean
government MOEHRD (R14-2002-059-010000-0).
\end{acknowledgments}

\end{document}